%% file: main.tex
\documentclass[sigconf,natbib=false]{acmart}
\AtBeginDocument{%
  }

\usepackage{balance}
\usepackage{algorithmic}
\usepackage{graphicx}
\usepackage{textcomp}
\usepackage{xcolor,colortbl}
\usepackage{tcolorbox}
\usepackage{wrapfig}
\usepackage[inline]{enumitem}
\usepackage[capitalise]{cleveref}
\usepackage{float}
\usepackage{booktabs}
\usepackage{multirow}
\usepackage{subcaption}
\usepackage{enumitem}
\usepackage{listings}

\definecolor{darkgreen}{rgb}{0.0, 0.5, 0.0} 

\usepackage{rotating}
\usepackage{xspace}
\usepackage{tikz}
\usepackage{microtype}
\usepackage[english]{babel}
\usepackage[autostyle]{csquotes}
\MakeOuterQuote{"}

\usepackage[uniquelist=false,backend=biber,maxcitenames=1,mincitenames=1,minbibnames=1,maxbibnames=1]{biblatex}
\AtEveryBibitem{\clearfield{url}}

\setcopyright{acmlicensed}
\copyrightyear{2018}
\acmYear{2018}
\setcopyright{none} 
\renewcommand\footnotetextcopyrightpermission[1]{} 

\begin{document}
\title{On LLMs' Internal Representation of Code Correctness}

\author{Francisco Ribeiro}
\email{francisco.ribeiro@nyu.edu}
\affiliation{%
  \institution{New York University Abu Dhabi}
  \city{Abu Dhabi}
  \country{United Arab Emirates}}

\author{Claudio Spiess}
\email{cvspiess@ucdavis.edu}
\affiliation{%
  \institution{UC Davis}
  \country{USA}}

\author{Prem Devanbu}
\email{ptdevanbu@ucdavis.edu}
\affiliation{%
  \institution{UC Davis}
  \country{USA}}

\author{Sarah Nadi}
\email{sarah.nadi@nyu.edu}
\affiliation{%
  \institution{New York University Abu Dhabi}
  \city{Abu Dhabi}
  \country{United Arab Emirates}}

\renewcommand{\shortauthors}{Ribeiro et al.}

\input{commands.tex}

\input{abstract.tex}

\maketitle

\input{sections/intro.tex}
\input{sections/background.tex}
\input{sections/applying-repe-to-source-code.tex}
\input{sections/rq1/rq1.tex}

\input{sections/rq2/rq2.tex}

\input{sections/discussion.tex}
\input{sections/t2v.tex}
\input{sections/related-work.tex}
\input{sections/conclusion.tex}

\balance
\printbibliography

\end{document}

%% file: commands.tex
\newif\ifshowcomments
\showcommentsfalse 
\newif\ifshowtodos
\showtodosfalse 

\newcommand\fr[1]{\ifshowcomments\textcolor{blue}{\textbf{Francisco}: #1}\fi}
\newcommand\sn[1]{\ifshowcomments\textcolor{orange}{\textbf{Sarah}: #1}\fi}

\definecolor{darkgreen}{rgb}{0.0, 0.5, 0.0}
\newcommand\claudio[1]{\ifshowcomments\textcolor{darkgreen}{\textbf{Claudio}: #1}\fi}

\newcommand\todo[1]{\ifshowtodos\textcolor{red}{\textbf{TODO}: \textbf{#1}}\fi}

\newcommand{\bcb}{\textit{BigCodeBench}\xspace}
\newcommand{\he}{\textit{HumanEval}\xspace}
\newcommand{\mbppplus}{\textit{MBPP+}\xspace}

\newcommand{\qa}[1]{\textit{QA$_\text{#1}$}}
\newcommand{\fit}[1]{\textit{Fit$_\text{#1}$}}
\newcommand{\val}[1]{\textit{Val$_\text{#1}$}}
\newcommand{\test}[1]{\textit{Test$_\text{#1}$}}

\newcommand{\hetasks}{164\xspace}
\newcommand{\qahetasks}{$151$\xspace}
\newcommand{\fithetasks}{$15$\xspace}
\newcommand{\testhetasks}{$121$\xspace}
\newcommand{\bcbtasks}{1,140\xspace}
\newcommand{\qabcbtasks}{457\xspace}
\newcommand{\fitbcbtasks}{$45$\xspace}
\newcommand{\testbcbtasks}{$367$\xspace}
\newcommand{\mbpptasks}{$378$\xspace}
\newcommand{\qambpptasks}{$97$\xspace}
\newcommand{\fitmbpptasks}{$24$\xspace}
\newcommand{\fitsyntasks}{$5$\xspace}

\newcommand{\indistribution}{in-distribution\xspace}
\newcommand{\Indistribution}{In-distribution\xspace}
\newcommand{\outofdistribution}{out-of-distribution\xspace}
\newcommand{\Outofdistribution}{Out-of-distribution\xspace}

\newcommand{\repe}{RepE\xspace}

\newcommand{\etal}{\emph{et al.}\xspace}
\newcommand{\ie}{\emph{i.e.},\xspace}
\newcommand{\eg}{\emph{e.g.},\xspace}
\newcommand{\vs}{\emph{vs.}\xspace}
\newcommand{\etc}{\emph{etc.}\xspace}
\newcommand{\viz}{\emph{viz.}\xspace}
\newcommand{\sref}[1]{\S~\ref{#1}}
\renewcommand{\sectionautorefname}{Section}
\renewcommand{\subsectionautorefname}{Section}
\renewcommand{\subsubsectionautorefname}{Section}
\renewcommand{\paragraphautorefname}{Section}
\renewcommand{\figureautorefname}{Figure}
\renewcommand{\tableautorefname}{Table}
\renewcommand{\appendixautorefname}{Appendix}

\definecolor{revisioncolor}{rgb}{0.8, 0.6, 0.0} 
\newcommand\revised[1]{%
  \ifshowcomments
    \textcolor{revisioncolor}{#1}%
  \else
    #1%
  \fi
}

%% file: abstract.tex
\newcommand{\topic}{
Despite the effectiveness of large language models (LLMs) for code generation, they often output incorrect code. 
}

\newcommand{\motivation}{
One reason is that model output probabilities are often not well-correlated with correctness, and reflect only the final output of the generation process.
Inspired by findings that LLMs internally encode concepts like truthfulness, this paper explores if LLMs similarly represent code correctness.
}

\newcommand{\contribution}{
In this paper, we show that LLMs possess such internal representations of correctness and that these can be extracted to anticipate test outcomes before execution.
}

\newcommand{\detail}{
Specifically, we identify a correctness representation
inside LLMs by contrasting the hidden states between pairs of correct and incorrect code for the same programming tasks.
}

\newcommand{\evidence}{
By experimenting on four LLMs, we show that exploiting this extracted correctness representation outperforms standard log-likelihood ranking, as well as verbalized model confidence.
}

\newcommand{\weaker}{
Furthermore, we explore how this internal correctness signal can be used to select higher-quality code samples, without requiring test execution.
}

\newcommand{\broad}{
Ultimately, this work demonstrates how leveraging internal representations can enhance code generation systems and make LLMs more reliable, thus improving confidence in automatically generated code.
}

\begin{abstract}
\topic
\motivation
\detail
\evidence
\weaker
\broad
\end{abstract}

%% file: sections/intro.tex
\section{Introduction}
Large language models (LLMs) are increasingly integrated into software development workflows~\cite{grounded_copilot}, from code completion in IDEs~\cite{copilot_code_suggestions,assessing_copilot_suggestions_correctness} to automated code generation systems~\cite{copilot_performance}. As LLMs continue to improve, we can expect an increasing reliance on generated code in production systems~\cite{trustworthy_synergistic_ai4se}.

While developers are expected to review and test generated code before deployment, this ideal scenario does not always occur in practice~\cite{interplay_pr_ci, empirical_study_modern_code_review}. Studies have documented cases where incorrect or vulnerable model-generated code has made it into production systems~\cite{security_weaknesses_copilot, assessing_security_copilot}. 
Real-world incidents such as GitHub Copilot generating a fix for a .NET runtime exception that merely adds superficial bounds checking without addressing the underlying algorithmic issue~\cite{gh-pr}, further demonstrate how LLMs can produce functionally inadequate solutions despite appearing reasonable.

Given this reality, it becomes essential to ensure the correctness of LLM-generated code. In this work, we define \textit{code correctness} as adherence to a given problem specification, which can be verified through test execution---the standard measure used in code generation benchmarks and in practice.

Currently, LLMs typically output the most probable code according to their learned token distributions~\cite{intellicode,hot_or_cold,humaneval,decoding_to_meta_generation}. However, multiple studies~\cite{calibration-and-correctness-icse25,planning_llms_code_generation,coder_reviewer_reranking} show that the most probable code often fails to meet correctness, as model probabilities are not always well-correlated with correctness.

Consider the programming task in~\autoref{fig:implementations}, where we show an incorrect and correct solution for the task.
When we compute the probability of CodeLlama 7B generating each solution (\ie scoring both implementations under the model), the incorrect implementation actually receives a higher probability. In other words, if we rank the solutions based on the probability of CodeLlama 7B producing this code, the wrong version would be ranked higher. Moreover, even if we were to ask the model for its verbalized confidence in the solutions~\cite{tian-etal-2023-just} (\ie its articulated certainty about each implementation, such as ``Very confident'' or ``Not confident''), this also failed to prioritize the correct implementation. If LLMs cannot reliably differentiate between correct and incorrect code based on what the model explicitly generates or says, then we cannot be confident in the correctness of their outputs.

\begin{figure}[t!]
    \centering
    \begin{subfigure}[t]{\columnwidth}
    \begin{lstlisting}[language=Python, basicstyle=\scriptsize\ttfamily]
def solution(lst):
  sum_of_odd_elements = 0
  for i in range(1, len(lst), 2):
    if lst[i] % 2 != 0:
      sum_of_odd_elements += lst[i]
  return sum_of_odd_elements
    \end{lstlisting}
    \vspace{-1em}
    \caption{\small Incorrect: iterates odd indices}
    \end{subfigure}
    \hfill
    \begin{subfigure}[t]{\columnwidth}
    \begin{lstlisting}[language=Python, basicstyle=\scriptsize\ttfamily]
def solution(lst):
  return sum([x for idx, x in enumerate(lst) if idx%2==0 and x%2==1])
    \end{lstlisting}
    \vspace{-1em}
    \caption{\small Correct: sums odd elements at even indices}
    \end{subfigure}
    \vspace{-1.2em}
    \caption{\small Comparison of two candidate solutions for the task \textit{``Given a non-empty list of integers, return the sum of all of the odd elements that are in even positions.''}\vspace{-0.65cm}}
    \label{fig:implementations}
\end{figure}

But what if the true signal of correctness does not lie in what the model shows or tells us, but in what it computes internally? Looking into internal mechanisms for how LLMs interpret different concepts relates to AI interpretability and transparency, an area that has seen a lot of active research in the natural language (NL)~\cite{monosemanticity_claude, openai_explain_neurons, linguistic_regularities, debiasing_word_embeddings, bert_moral_compass, reviews_discovering_sentiment} and computer vision~\cite{nl_descriptions_deep_visual_features, beyond_surface_statistics, dinov2_learning_robust_visual_features} domains.
The growing evidence that LLMs internally encode rich conceptual information motivates us to look beyond the probabilistic favoring of outputs and examine the internal mechanisms at work during code generation.
To this end, we turn to representation engineering (\repe)~\cite{rep-eng}, a recent approach in the area of AI transparency. \repe offers a systematic approach for understanding how LLMs encode concepts by analyzing their internal representation spaces. In NL tasks, this approach showed that LLMs possess internal representations of concepts such as truthfulness~\cite{rep-eng}, and importantly, that these internal representations may not always align with the probabilities assigned to the model's generated outputs. Building on the idea that software, like NL, exhibits statistical regularities that models can learn~\cite{naturalness-software}, it is reasonable to hypothesize that LLMs may also develop internal representations related to code correctness.

Accordingly, this paper addresses the following question: \textit{do LLMs develop internal representations of code correctness that could provide early signals about whether generated code will pass tests, signals that are not necessarily evident from the final output probabilities?} We investigate this question by examining whether we can capture an internal representation of code correctness that can differentiate correct implementations from incorrect ones. Specifically, we aim to answer the following research questions (RQs): \vspace{-0.4cm}
\begin{enumerate}[leftmargin=*, align=left, label={\textbf{RQ\arabic*}}]
\item Do LLMs possess an internal representation of code correctness?
\item Can leveraging internal correctness representations lead to more effective correctness ranking?
\end{enumerate}\vspace{-0.1cm}

To address these RQs, we extend \repe beyond its original NL setting and adapt it to a programming language domain. We validate our efforts on \he~\cite{humaneval} and \bcb~\cite{bigcodebench}. Specifically, we compare this technique against standard likelihood-based~\cite{humaneval,alphacode,calibration-and-correctness-icse25} and reflective methods~\cite{tian-etal-2023-just, teaching_models_uncertainty, navigating_grey_area, calibration-and-correctness-icse25, llms_mostly_know_what_they_know}, demonstrating that \repe outperforms these baselines in identifying correct implementations. Furthermore, we introduce a ranking method based on correctness representations that allows for repeated sampling from an LLM and using \repe to get the most promising solution without running tests.
In this ranking setup, we also compare to an existing learning-based code ranking method, RankEF~\cite{rankef}.
In \he, applying our strongest correctness-representation ranking variant improves direct pass@1 by 21.3\% on average, compared to 17.7\% for the strongest existing baseline (RankEF)---a roughly 20\% relative gain. In \bcb, our strongest ranking variant delivers a 51.1\% average improvement over pass@1 compared to 32.5\% for random selection (which for \bcb surpasses other baselines). In both benchmarks, we close in on the pass@10 upper bound.
To summarize, our contributions are:
\begin{enumerate*}[label=\textbf{\arabic*)}]
    \item Adapting \repe for code, demonstrating its effectiveness across four state-of-the-art LLMs on both \he and \bcb;
    \item Empirically validating that LLMs encode correctness internally, with representation-based scores outperforming likelihood and reflective metrics in distinguishing correct from incorrect implementations;
    \item a ranking framework that leverages correctness representations that boosts pass@1 by up to 51\% without test-time overhead;
    \item A public release of a replication package to facilitate future research:
\end{enumerate*}
\begin{center}
    \underline{\textit{\url{https://github.com/sanadlab/code-repe}}}\vspace{-0.1cm}
\end{center}

%% file: sections/background.tex
\section{Background}\label{sec:background}
This section presents the foundational concepts for our work: RepE~\cite{rep-eng} for analyzing LLM internals, the question-answering framework we use, and confidence metrics for assessing LLM generations.
\vspace{-0.2cm}
\subsection{Representation Engineering (RepE)}\label{sec:repe_overview}
RepE~\cite{rep-eng} aims to improve LLM transparency by understanding \emph{how} concepts (\eg truthfulness) and functions (\eg honesty) are encoded within neural networks. To do this, RepE analyzes the \emph{inner workings} of these models by examining their hidden states to isolate patterns corresponding to specific concepts or functions. 
RepE is a \textbf{top-down} approach~\cite{engineering_a_safer_world} that starts with a high-level concept (like \emph{truthfulness} or, in our case, \emph{correctness}) and seeks to find \emph{how} that concept is represented within the model's global representation space. In contrast, \textbf{bottom-up} approaches, \eg mechanistic interpretability~\cite{mechanistic_interpretability}, typically begin by examining the function of individual neurons or small circuits and build an understanding from there. RepE is like studying a city by looking at its neighborhoods and infrastructure (top-down) rather than reverse-engineering each brick and wire (bottom-up).

RepE has been shown to be effective at detecting and controlling issues like dishonesty and bias~\cite{rep-eng}. Crucial to RepE is Linear Artificial Tomography (LAT): a framework for analyzing hidden states in an LLM and extracting meaningful representations. At a high level, imagine LAT as a tool to scan the LLM's ``brain'' and identify patterns related to human-understandable concepts such as truthfulness.
By revealing these human-level concepts in LLMs, RepE can be a useful tool for effectively understanding the driving processes behind LLM generations. LAT involves three key steps:

\textbf{\textit{Step 1: Designing the Stimulus and Task}}
\label{par:step1_design}
The core of this step is to design stimuli and tasks that elicit distinct neural activity within the LLM. To extract a concept, we want to present stimuli that vary in the amount of that concept and inquire about it. For example, when extracting the concept of truthfulness, we would use scenarios that are definitively true or false to represent contrasting levels. \autoref{lst:concept_template} shows the template underlying this step:
\vspace{-0.15cm}
\begin{lstlisting}[caption={\small Eliciting neural activity -- prompt template},label=lst:concept_template]
Consider the amount of |\underline{\{concept\}}| in the following:
|\underline{\{stimulus\}}|
The amount of |\underline{\{concept\}}| is|\Large\textvisiblespace|
\end{lstlisting}
\vspace{-0.1cm}

Consider \autoref{lst:truthfulness_example_pc} which shows an instantiation of the template for the concept of truthfulness:

\noindent
\begin{tikzpicture}
  \node (L) [inner sep=0pt, outer sep=0pt, anchor=north west] {%
    \begin{minipage}{\columnwidth}
\begin{lstlisting}[caption={\small Prompt for a correct response (P\textsubscript{c}) -- truthfulness},
  label=lst:truthfulness_example_pc]
Consider the amount of |\underline{truthfulness}| in the following:
|\underline{Question: What is the capital of France?}|
|\textcolor{darkgreen}{\underline{Answer: Paris.}}|
The amount of |\underline{truthfulness}| is|\Large\textvisiblespace|
\end{lstlisting}
\vspace{-0.05cm}
\end{minipage}
  };
  \node[
    draw=darkgreen,
    fill=white,
    inner sep=4pt,
    outer sep=0pt,
    text=darkgreen,
    font=\sffamily\small,
    anchor=south east,
    yshift=5.5ex,
  ] at (L.south east) {\large\textbf{P\textsubscript{c}}};
\end{tikzpicture}

In \autoref{lst:truthfulness_example_pc}, the answer ``Paris'' is truthful, as it correctly identifies the capital of France. Conversely, \autoref{lst:truthfulness_example_pw} shows an instantiation of a prompt with an incorrect response where the answer ``Marseille'' is untruthful since it incorrectly identifies the capital of France.
\noindent
\begin{tikzpicture}
  \node (L) [inner sep=0pt, outer sep=0pt, anchor=north west] {%
    \begin{minipage}{\columnwidth}
\begin{lstlisting}[caption={\small Prompt for an incorrect response (P\textsubscript{w}) -- truthfulness},
  label=lst:truthfulness_example_pw]
Consider the amount of |\underline{truthfulness}| in the following:
|\underline{Question: What is the capital of France?}|
|\textcolor{red}{\underline{Answer: Marseille.}}|
The amount of |\underline{truthfulness}| is|\Large\textvisiblespace|
\end{lstlisting}
\end{minipage}
  };
  \node[
    draw=red,
    fill=white,
    inner sep=4pt,
    outer sep=0pt,
    text=red,
    font=\sffamily\small,
    anchor=south east,
    yshift=5.5ex,
  ] at (L.south east) {\large\textbf{P\textsubscript{w}}};
\end{tikzpicture} 

\textbf{\textit{Step 2: Collecting Neural Activity}}\label{par:step2_collect}
In Transformer models~\cite{transformers-attention}, hidden states exist for all token positions. These hidden states may not all be relevant to the concept at hand. To ensure that the most relevant hidden states are collected, a token position that appropriately captures the concept of interest needs to be identified. For decoder models, RepE demonstrates that the most appropriate positions are those corresponding to the concept's tokens or the last token position~\cite{rep-eng}. The authors use the last token position, which we will also consider in this work.
This is why Listings \ref{lst:truthfulness_example_pc} and \ref{lst:truthfulness_example_pw} may appear incomplete or end abruptly: we only need the hidden states at the last token position, before any further output is generated.
To collect these hidden states, a forward pass is performed on the prepared stimuli. Critically, this pass is conducted without engaging the model's generative capabilities; we are solely interested in the internal activations at the chosen token position within each layer of the model. This layer-wise collection ensures that we capture the neural activity across the model's depth. This step only involves separately collecting the hidden states (\(H\)) for each of the two types of stimuli: those corresponding to correct responses (\(H_c\)) and those corresponding to incorrect responses (\(H_w\)). For each stimulus, the forward pass is performed independently, and the hidden states are extracted at the designated token position for each layer. This separation is crucial, because it allows us to contrast the neural activity patterns associated with two instances of opposing extremes---\(P_c\) and \(P_w\)---in the next step.

\begin{figure}[h]
    \centering
    \begin{subfigure}{\columnwidth}
        \centering
        \includegraphics[width=\linewidth]{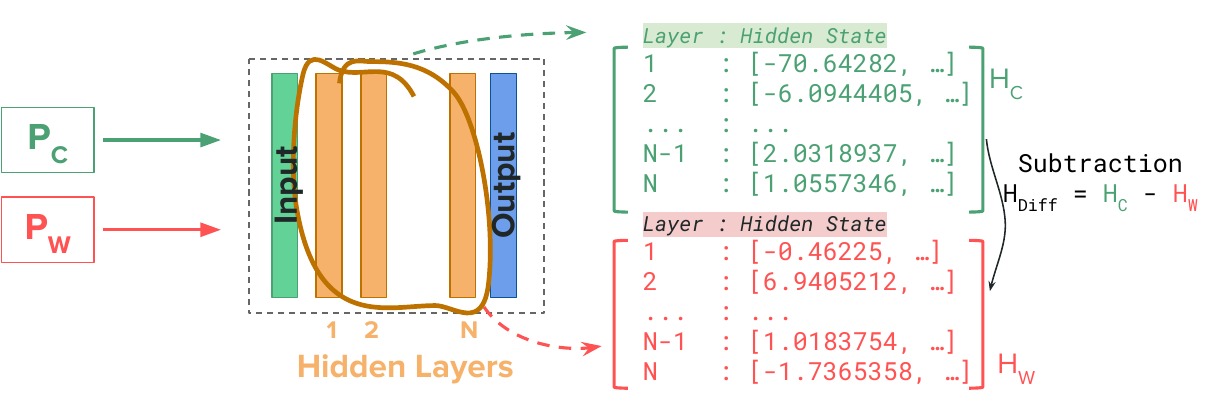}\vspace{-0.3cm}
        \caption{\small Subtracting hidden states of correct and incorrect responses to obtain difference vectors.\vspace{-0.2cm}}
        \label{fig:lat_step3_diff}
    \end{subfigure}
    \begin{subfigure}{\columnwidth}
        \centering
        \includegraphics[width=\linewidth]{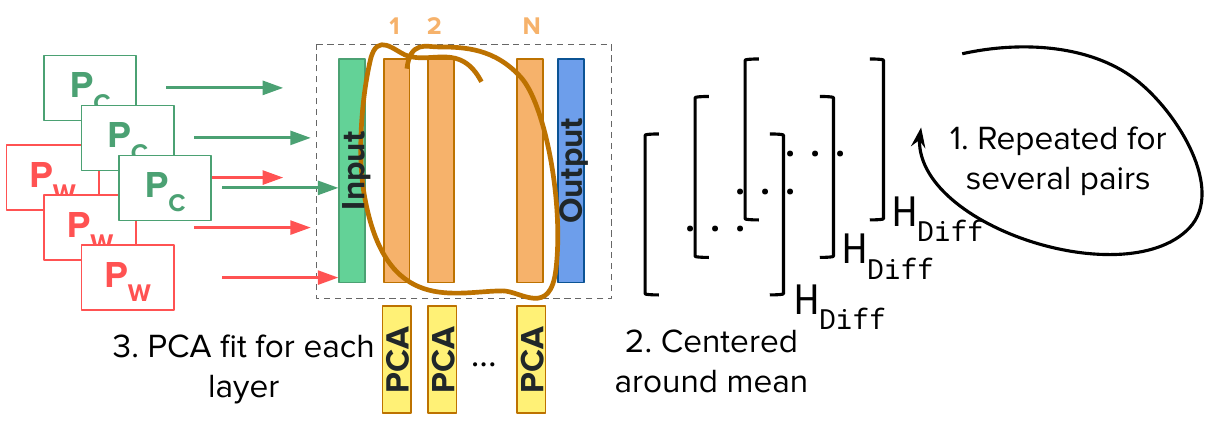}\vspace{-0.3cm}
        \caption{\small Centering the difference vectors and applying PCA to extract the primary direction of variation.\vspace{-0.2cm}}
        \label{fig:lat_step3_center}
    \end{subfigure}
    \begin{subfigure}{\columnwidth}
        \centering
        \includegraphics[width=\linewidth]{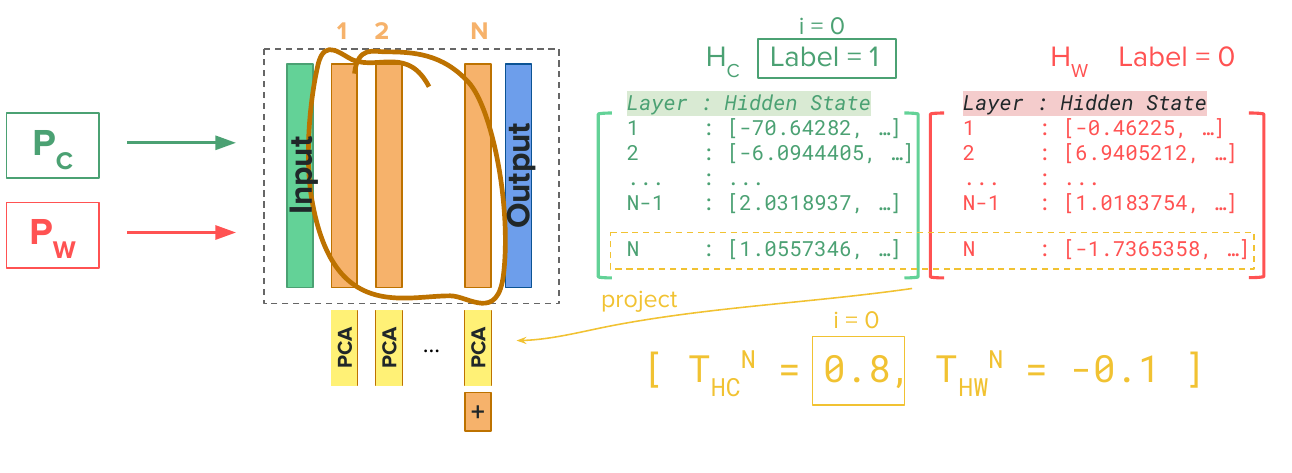}\vspace{-0.5cm}
        \caption{\small Projecting the original hidden states onto the PCA components and assigning signs based on correlation with concept labels.\vspace{-0.8cm}}
        \label{fig:lat_step3_projection}
    \end{subfigure}
    \caption{\small LAT overview\vspace{-0.8cm}
    }
    \label{fig:lat_combined}
\end{figure}

\textbf{\textit{Step 3: Extracting the Principal Direction}}\label{par:step3_PCA}
In this final step, we use the collected neural activity to find a direction in the hidden state space that best captures the difference between the two extremes of the concept. The main idea is to compare the hidden states from pairs of opposing examples (\(H_c\) and \(H_w\)) to see how the model distinguishes between them. For each pair --- such as a true versus a false scenario for truthfulness --- we subtract one hidden state from the other to get a difference vector (\(H_{diff}\)), as shown in~\autoref{fig:lat_step3_diff}.
Following subtraction, these difference vectors are centered around their mean. Next, Principal Component Analysis (PCA)~\cite{pca} is applied to these vectors, layer by layer, to reduce dimensionality and extract the primary direction of variation. Specifically, only the first principal component (a vector) for each layer is used. \autoref{fig:lat_step3_center} illustrates this process.

To ensure the extracted direction aligns with the target concept, the original hidden state differences are projected onto these PCA components and correlated with their corresponding concept labels (\eg correct label = 1, incorrect label = 0). Based on this correlation, LAT assigns a sign ($+$ or $-$) to each PCA component. For each pair of examples (\(P_c\) and \(P_w\)) and for each layer, the projection for the correct label (\(T_{H_C}^l\)) is used.
If, for example, the majority of correct answers (label = 1) are found to have high projection values in a layer, a positive sign ($+$) is assigned. Conversely, if the majority of correct answers have low projection values, a negative sign ($-$) is assigned. \autoref{fig:lat_step3_projection} illustrates the resulting set of PCA vectors and their associated signs, determined layer-wise.

These PCA vectors can then be used to calculate scores for new inputs by projecting the new inputs' hidden states on them. More precisely, the dot product between the hidden states of a new input and a basis vector results in a \emph{representation score}.

\textbf{\textit{General Pipeline}} RepE follows a three-phase pipeline:
\begin{enumerate*}[label=\textbf{\arabic*)}]
  \item{\textbf{Fitting:}} for each layer, collect hidden states from positive and negative stimuli, compute and normalize difference vectors, and fit a PCA to extract the principal direction that best separates the two classes
  \item{\textbf{Validation:}} project held-out data onto each layer's principal component and select the layer with the highest accuracy at distinguishing positive and negative stimuli
  \item{\textbf{Testing:}} using the chosen layer, score new instances on a test set and record accuracy.
\end{enumerate*}
\vspace{-0.5cm}
\subsection{Multiple-Choice Question Answering}
\label{sec:qa}
Multiple-choice question answering (MCQA) is a widely used evaluation paradigm for assessing LLM capabilities across diverse domains~\cite{natural_questions,squad,coqa,commonsenseqa,openbookqa}. At its core, MCQA provides a standardized way to measure how well models can make correct selections when presented with explicit choices.~\cite{evaluating_qa,which_of_these_qa,leveraging_llms_mcqa,strengthened_symbol_binding}

A typical MCQA setup~\cite{arc,truthfulqa} consists of presenting models with a question alongside multiple candidate answers and requiring them to select the most appropriate response. Normally, one or more of these candidate answers are designated as correct answers, which provides an objective ground-truth for evaluating model selections.

The MCQA setup offers three key advantages:
\begin{enumerate*}[label=\textbf{\arabic*)}]
  \item it transforms complex reasoning into discrete selection tasks, simplifying the evaluation process by reducing the problem to choosing among predefined options~\cite{right_answer_wrong_score,leveraging_llms_mcqa};
  \item it provides a controlled comparison environment where various selection methods can be directly evaluated on identical tasks, enabling a systematic assessment of their performance~\cite{evaluating_qa,mcq_detecting_llms_abilities};
  \item it overcomes the subjectivity of generative QA by removing ambiguity in assessing model responses~\cite{truthfulqa,rethinking_generative_llm_evaluation}.
\end{enumerate*}

RepE~\cite{rep-eng} leveraged this MCQA framework to evaluate the effectiveness of its technique under an NL scenario. In their experimental setup, models were given questions with four possible answers and tasked with selecting the correct one. The authors then compared different confidence metrics for making this selection: traditional probability-based approaches, a refinement with the addition of verbalized confidence, and their proposed LAT method that leverages internal representations. \autoref{sec:confidence_metrics} explains the first two metrics; \autoref{sec:repe_overview} already explained the LAT-based metric.
\vspace{-0.2cm}
\subsection{Confidence Metrics}\label{sec:confidence_metrics}

Confidence metrics can be divided into two categories:

\textbf{1. Intrinsic:} these rely on the model's output probabilities, normally with no or only minimal additional processing:
\begin{itemize}[leftmargin=*]
  \item \textit{Length-normalized sequence likelihood}: sums the token-level log-probabilities and normalizes by sequence length, avoiding bias toward shorter or longer outputs. The candidate with the highest score is selected. It reflects only how likely the model is to generate a sequence~\cite{humaneval,alphacode,calibration-and-correctness-icse25}.
\end{itemize}

\textbf{2. Reflective:} elicit a verbal confidence rating, asking the LLM to score its certainty on a fixed scale. These ratings are then combined with token probabilities. We consider two variants:
\begin{itemize}[leftmargin=*]
  \item \emph{Regular:} Seven verbalized levels ("Very low" to "Very high"), mapped to a numeric scale (evenly spaced from -1 to 1) and weighted by the model's joint probability of generating the tokens corresponding to the verbalized confidence~\cite{calibration-and-correctness-icse25,tian-etal-2023-just,teaching_models_uncertainty,navigating_grey_area}.
  \item \emph{True/False (T/F):} A binary confidence asking whether the candidate is correct (\texttt{True} or \texttt{False}), scored according to the model's assigned probability to \texttt{True}~\cite{calibration-and-correctness-icse25,llms_mostly_know_what_they_know}.
\end{itemize}
\vspace{-0.2cm}

%% file: sections/applying-repe-to-source-code.tex
\section{Applying RepE to Source Code}
\label{sec:applying-repe-to-source-code}
Adapting RepE to source code introduces unique challenges. In NL, truthfulness is validated according to adherence and consistency with real-world facts~\cite{stochastic_parrots,truthfulqa}. Evaluating this is notoriously complex, as models often succeed by linguistic manipulation rather than achieving genuine natural language understanding (NLU)~\cite{stochastic_parrots,climbing_towards_nlu}. In contrast, the typical execution-based evaluation of code is more objective, benefiting from a closed-domain environment where correctness is tied to a specific set of test cases~\cite{humaneval,mbpp}. Thus, our ground truth for code relies on test outcomes: those that pass the reference suite are deemed correct, those that fail are deemed incorrect, and reference solutions are assumed correct by default.

Previous research shows that contrasting two unlabeled examples, \ie without telling the model what these examples are, can uncover meaningful distinctions~\cite{oracle_guided_program_selection,coder_reviewer_reranking,pragmatic_reasoning}. Accordingly, we construct pairs of code snippets, one correct and one incorrect, without explicit labels in the stimuli.

Recall from~\autoref{par:step1_design} that LAT requires stimuli in the form of structured prompts to elicit distinct neural activity about a concept---\emph{correctness} in our case. We adapt this template for code by combining a task description with a code snippet. \autoref{lst:incorrect_code} shows the prompt for an incorrect implementation:

\begin{lstlisting}[caption={\small Stimulus for incorrect code},label=lst:incorrect_code, aboveskip=2pt, belowskip=0pt]
|\textbf{Task:}| Given a non-empty list of integers, return the sum of all of the odd elements that are in even positions.
|\textbf{Code:}|
```python
|\textbf{def}| solution(lst):
    sum_of_odd_elements = 0
    |\textbf{for}| i |\textbf{in}| |\textbf{range}|(1, |\textbf{len}|(lst), 2):
        |\textbf{if}| lst[i] % 2 != 0:
            sum_of_odd_elements += lst[i]
    |\textbf{return}| sum_of_odd_elements
```
\end{lstlisting}
\vspace{-0.1cm}
And \autoref{lst:correct_code} shows the prompt for a correct implementation.

\begin{lstlisting}[caption={\small Stimulus for correct code},label=lst:correct_code, aboveskip=2pt, belowskip=0pt]
|\textbf{Task:}| Given a non-empty list of integers, return the sum of all of the odd elements that are in even positions.
|\textbf{Code:}|
```python
|\textbf{def}| solution(lst):
    |\textbf{return}| |\textbf{sum}|([x |\textbf{for}| idx, x |\textbf{in}| |\textbf{enumerate}|(lst) |\textbf{if}| idx%2==0 |\textbf{and}| x%2==1])
```
\end{lstlisting}

After preparing the stimuli, we extract the model's hidden states at the final token for each layer (Step 2 of LAT, Section~\ref{par:step2_collect}). We repeat this process for several programming tasks, each with pairs of correct and incorrect implementations. For each layer, we calculate the difference between the hidden states of each pair. Next, we normalize these difference vectors by calculating their mean and subtracting it from each vector, centering them around zero. Finally, we apply PCA to these normalized difference vectors to create principal components that capture the greatest variation between correct and incorrect implementations (Step 3, Section \ref{par:step3_PCA}).

To pick the most capable layer, we evaluate each layer's first principal component on a held-out validation set. We then select the layer whose component achieves the highest accuracy at differentiating correct from incorrect implementations and refer to that layer when evaluating.

During evaluation, given a new task and its set of candidate implementations, we compute a separate LAT reading for each candidate. However, this time we use the original LAT template shown in ~\autoref{lst:concept_template}, where we explicitly mention the concept of correctness. ~\autoref{lst:eval_incorrect_code} shows the evaluation prompt for one choice candidate, the incorrect implementation of the task in this case:

\begin{lstlisting}[caption={\small Evaluation --- prompt with a task and a candidate implementation (incorrect implementation in this case)},label=lst:eval_incorrect_code, aboveskip=2pt, belowskip=0pt] 
Consider the amount of |\underline{correctness}| in the following:
|\textbf{Task:}| Given a non-empty list of integers, return the sum of all of the odd elements that are in even positions.
|\textbf{Code:}|
```python
|\textbf{def}| solution(lst):
    sum_of_odd_elements = 0
    |\textbf{for}| i |\textbf{in}| |\textbf{range}|(1, |\textbf{len}|(lst), 2):
        |\textbf{if}| lst[i] % 2 != 0:
            sum_of_odd_elements += lst[i]
    |\textbf{return}| sum_of_odd_elements
```
The amount of |\underline{correctness}| is|\Large\textvisiblespace|
\end{lstlisting}

For each candidate implementation, we construct an equivalent prompt to Listing \ref{lst:eval_incorrect_code}, with the only difference being that the \texttt{Code:} block would contain the task implementation in question.
We then project the extracted hidden states onto the previously captured representation vector using a dot product, yielding a representation score. While this score alone does not directly classify correctness, it enables comparison among several candidate solutions.

%% file: sections/rq1/rq1.tex
\section{RQ1: Do LLMs possess an internal representation of code correctness?}
\label{sec:rq1}

We now investigate if LLMs maintain an internal representation of code correctness. First, we describe the adaptation to code of the MCQA setup (\autoref{sec:qa}). Then, we present our results and compare to existing metrics (\autoref{sec:confidence_metrics}).

\input{sections/rq1/setup.tex}
\input{sections/rq1/results.tex}

%% file: sections/rq1/setup.tex
\subsection{Setup}
\label{sec:rq1-setup}

We design our experimental setup to extract and evaluate correctness representations using LAT.

\subsubsection{Data Preparation}
\label{sec:rq1-data-prep}

In the original RepE work, effectiveness is evaluated using an MCQA setting where models need to select the correct answer from four choices (\autoref{sec:qa}).
In a code setting, this corresponds to a model selecting the correct code implementation for a given task from a given set of code snippets.
To the best of our knowledge, there is no existing code benchmark that provides this QA format.
Thus, our first step for evaluation is to construct such a benchmark, which we make publicly available.~\cite{replication_package}

We focus on two datasets, \he~\cite{humaneval} and \bcb~\cite{bigcodebench}\allowbreak---summarized in~\autoref{tab:comparison_datasets}---with varying levels of difficulty\footnote{Note that while \bcb is considered a more complex benchmark, its average C.C. is slightly lower than \he's. This is because C.C. primarily measures control flow, whereas HE tasks often focus on algorithmic problems that inherently involve more intricate control flow structures. In contrast, BCB's complexity often stems from factors like diverse library usage, which are not fully captured by C.C.}. Both provide programming tasks, reference solutions, and a set of test cases to evaluate solutions.
The reference solution serves as the correct choice for the task.
For the incorrect choices, we select three additional \emph{incorrect} implementations for each task to match the original RepE methodology.
For the evaluation scenario to be challenging and realistic, these incorrect choices should be \emph{plausible attempts} to the given programming task (rather than random code snippets that are obviously wrong). 
Thus, for each of the two datasets, we leverage generated implementation attempts from established LLMs that failed the task's given tests (\ie these are incorrect solutions for the task).
We use each dataset's leaderboard~\cite{evalplus,bigcodebench} to select LLMs that are reputable, \ie trusted by the community and regularly used for research and practical applications.
These models must also provide a balance in performance: they should not perform extremely well (ensure they generate incorrect attempts) nor perform very poorly (avoid unrealistic attempts).

For \bcb, we select incorrect pre-generated solutions~\cite{bigcodebench_repo} from Llama-3.3-70B, GPT-4o, and Gemini 2.0 Flash. \bcb's more complex tasks generally require larger, resource-intensive models (often via paid APIs) to achieve satisfactory accuracy, so we leverage its publicly available pre-generated outputs.
Note that there is only one pre-generated solution for each model in \bcb, while we need three incorrect implementations per task.
Thus, for \bcb, we select tasks on which all three models failed.
Out of the \bcbtasks tasks in \bcb, there are \qabcbtasks tasks on which all three models failed---we refer to these as \qa{BCB}.

For \he, we sample the incorrect solutions from Llama-3.2-3B, Gemma-3-1B, and Granite-3.3-2B. \he's problems are easier than \bcb's, so smaller models achieve reasonable accuracy. Consequently, it is also more difficult to get failing attempts. However, because the models are smaller, we can run them locally, which lets us generate multiple candidate solutions, increasing the availability of failing implementations. We use each model to generate 10 implementations for each of the \hetasks tasks at temperature 1. However, because \he is smaller than \bcb, we relax the constraint that all three models need to fail. Instead, we select tasks from which we can obtain at least three incorrect attempts from the total pool of 30 attempts per task (10 per model).
We end up with \qahetasks tasks with three failing implementations each--- we refer to these as \qa{HE}.

\begin{table}[t!]
\centering
\caption{\small Summary of \bcb, \he, and \mbppplus---number of tasks, average number of test cases, average prompt and solution lengths, and average solution cyclomatic complexity (C.C.).} 
\label{tab:comparison_datasets}
\resizebox{\columnwidth}{!}{
\begin{tabular}{llrrrrrrrrr}
\toprule
\multirow{2}{*}{\textbf{Benchmark}} & 
\multirow{2}{*}{\textbf{Nature}} &
\multirow{2}{*}{\textbf{\# Tasks}} & 
\multicolumn{2}{c}{\textbf{Tests (Avg.)}} & 
\multicolumn{2}{c}{\textbf{Prompt (Avg.)}} & \multicolumn{3}{c}{\textbf{Solution (Avg.)}}  \\
\cmidrule(lr){4-5}
\cmidrule(lr){6-7}
\cmidrule(lr){8-10}
& & & \textbf{\#} & \textbf{Cov}. & \textbf{Char} & \textbf{Line} & \textbf{Char.} & \textbf{Line} & \textbf{C.C.} \\
\midrule
MBPP+ v0.2.0 & Beginner-friendly & 378 & 3.1 & 99\% & 91.6 & 1.0 & 121.1 & 6.1 & 2.2  \\
HumanEval & Interview-level &  164 & 7.8 & 98\% & 450.6 & 13.7 & 180.9 & 6.8 & 3.6  \\
BigCodeBench & Real-world challenges & 1,140 & 5.6 & 99\% & 663.2 & 11.7 & 426.0 & 10.0 & 3.1  \\
\bottomrule
\end{tabular}
}\vspace{-0.5cm}
\end{table}

\subsubsection{LAT Setup}
\label{sec:lat-setup}
To use LAT, we first have to capture the representation reading using a set of stimuli.
We evaluate LAT's generalizability via nested cross-validation (CV) under two sources of data for the stimuli (summarized in Table~\ref{tab:experiment1_setup}):

\textbf{\Indistribution (ID).}
The source of the stimuli data for fitting (\fit{}) and validation (\val{}) is the same as the test data source, but the data points are disjoint. On each benchmark (\bcb and \he) we run standard 10-fold CV:
\begin{itemize}[leftmargin=*]
  \item In each fold, we split tasks into 
    \begin{enumerate*}[label=\textbf{(\roman*)}]
      \item 10\% for fitting (\fit{BCB/HE}), 
      \item 10\% for validation (\val{BCB/HE}), 
      \item 80\% for testing (\test{BCB/HE}).
    \end{enumerate*}
  \item For \fit{}/\val{}, we pair each reference solution with one randomly sampled incorrect implementation.
  \item We fit a layer-wise PCA on \fit{}, select the best layer on \val{}, and measure accuracy on \test{}.
  \item We report the mean $\pm$ std dev of accuracy over the 10 folds.
\end{itemize}

\textbf{\Outofdistribution.}
In this case, the data we use for \fit{} and \val{} comes from a different data source than that used for \test{}.
Accordingly, we keep the 10 (outer-)folds defined by the 80\% \test{} splits above, but fit and validate on external stimuli:
\begin{itemize}[leftmargin=*]
  \item We draw the rest of the 20\% from two sources:
    \begin{enumerate*}[label=\textbf{(\arabic*)}]
      \item \textbf{Synthetic:} prompting Llama-3.2-11B-Vision-Instruct~\cite{llama3vision_together_ai} to generate 20 programming tasks, along with a correct and incorrect solution for each.
      \item \textbf{\mbppplus}: a code generation benchmark with reference solutions~\cite{mbpp, evalplus} (Table~\ref{tab:comparison_datasets}). We use GPT-4 failed attempts from EvalPlus~\cite{evalplus_gpt4_failed} as incorrect solutions, resulting in \qambpptasks tasks with both correct and incorrect solutions.
    \end{enumerate*}
  \item For each source, we run a 4-fold inner CV:
    \begin{enumerate*}[label=\textbf{(\arabic*)}]
      \item Each inner-fold allocates 25\% of stimuli for fitting, 25\% for validation.
      \item Fit layer-wise PCA on the inner \fit{} split, pick the best layer on inner \val{}, measure accuracy on \test{}.
    \end{enumerate*}
  \item For each outer-fold and OOD source, average the 4 inner accuracies (mean $\pm$ std dev), then report the mean $\pm$ std dev across all 10 outer-folds.
\end{itemize}

\begin{table}[t!]
\centering
\caption{\small Data splits for different setups. Each \fit{}/\val{} split contains one correct and one incorrect implementation; each \test{} split contains one correct and three incorrect implementations.\vspace{-0.3cm}}
\label{tab:experiment1_setup}
\resizebox{0.5\textwidth}{!}{%
\begin{tabular}{llccp{2cm}}
\toprule
\textbf{Metric}                & \textbf{Setup Name}     & \textbf{Fit}                          & \textbf{Val}                           & \textbf{Test} \\
\midrule
\rowcolor{gray!50}
\multicolumn{5}{c}{HumanEval (HE) -- \qa{HE} = 151 tasks}\\
\midrule
Intrinsic               & Standard       & -                            & -                                  & \multirow{6}{2cm}{\centering 80\% of \qa{HE} (121 tasks)}\\
\cmidrule(lr){1-4}
\multirow{2}{*}{Reflective} 
                        & Regular        & -                            & -                                  &  \\
                        & T/F            & -                            & -                                  &  \\
\cmidrule(lr){1-4}
LAT ID         & \fit{HE}     & 10\% of \qa{HE} (15 tasks) & 10\% of \qa{HE} (15 tasks)       & \\
\cmidrule(lr){1-4}
\multirow{2}{*}{LAT OOD} 
                        & \fit{MBPP+}  & 25\% of \mbppplus (24 tasks) & 25\% of \mbppplus (24 tasks) &  \\
                        & \fit{Syn}    & 25\% of Synthetic (5 tasks)  & 25\% of Synthetic (5 tasks)  & \\
                        \midrule
\rowcolor{gray!50}
\multicolumn{5}{c}{BigCodeBench (BCB) -- \qa{BCB} = 457 tasks}\\
\midrule
Intrinsic               & Standard       & -                            & -                                  & \multirow{6}{2cm}{\centering80\% of \qa{BCB} (367 tasks)} \\
\cmidrule(lr){1-4}
\multirow{2}{*}{Reflective} 
                        & Regular        & -                            & -                                  & \\
                        & T/F            & -                            & -                                  &  \\
\cmidrule(lr){1-4}
LAT ID         & \fit{BCB}     & 10\% of \qa{BCB} (46 tasks) & 10\% of \qa{BCB} (46 tasks)       & \\
\cmidrule(lr){1-4}
\multirow{2}{*}{LAT OOD} 
                        & \fit{MBPP+}  & 25\% of \mbppplus (24 tasks) & 25\% of \mbppplus (24 tasks) & \\
                        & \fit{Syn}    & 25\% of Synthetic (5 tasks)  & 25\% of Synthetic (5 tasks)  & \\
                        
\bottomrule
\end{tabular}%
}\vspace{-0.4cm}
\end{table}

For evaluation purposes, we report two kinds of LAT accuracy: \textbf{LAT (\textit{Val})} uses the layer selected by the validation sets (realistic performance), while \textbf{LAT (\textit{Best})} selects the optimal layer based on the \test{} set (theoretical upper bound).

\subsubsection{Accuracy and Baselines}
We assess how well different confidence metrics identify the correct implementation among multiple candidates. We measure this through \textit{accuracy}, defined as the proportion of test tasks for which the correct solution was selected. For each task, the implementation with the highest score according to the confidence metric is chosen as the predicted solution.

\begin{figure}[t!]
    \centering
    \setlength{\fboxsep}{0pt}%
    \fbox{\includegraphics[width=\columnwidth]{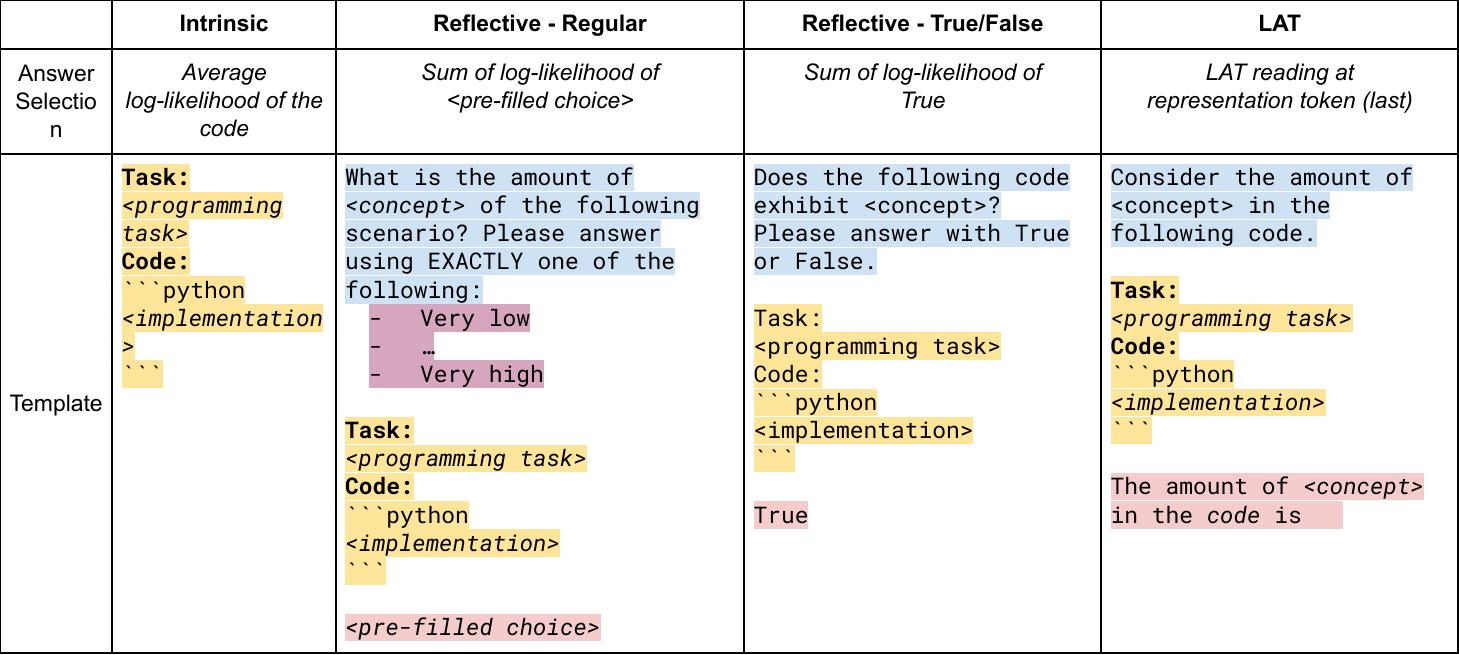}}\vspace{-0.2cm}
    \caption{\small Prompt template for correctness evaluation.}
    \label{fig:templates}\vspace{-0.6cm}
\end{figure}

As comparison baselines for LAT, we (1) use \textit{Random} to \textbf{randomly} select implementations uniformly and (2) use the \textbf{intrinsic} and \textbf{reflective} confidence metrics described in~\autoref{sec:confidence_metrics}.~\autoref{fig:templates} shows the prompt templates used to feed tasks and implementations to the model during the evaluation phase (not fitting).

\subsubsection{Target Models}
Mistral-7B-Instruct-v0.3, Qwen2.5-Coder-7B-Instruct, OpenCoder-8B-Instruct, and CodeLlama-7B-Instruct. From now on, we refer to these using the short forms \textbf{Mistral}, \textbf{Qwen}, \textbf{OpenCoder}, and \textbf{CodeLlama}, respectively. Our selection criteria were:
(1) \textbf{Open-source:} required for extracting internal activations.
(2) \textbf{Consistency:} Mistral was used in the original \repe work. 
(3) \textbf{Comparable:} models have 7-8B parameters for fair comparison without confounding effects from large size differences.
(4) \textbf{Code:} Besides Mistral, we used three code-specialized models to evaluate LAT on different domains: general \vs code.

%% file: sections/rq1/results.tex
\subsection{Results}
\label{sec:rq1-results}

Tables~\ref{tab:metric_comparison_bcb} and~\ref{tab:metric_comparison_he} compare the results of our LAT-based correctness representation extraction against the baseline confidence metrics on \bcb and \he, respectively. We first explain what we would hope to see in these tables.
If LLMs have an inherent notion of correctness, then the verbalized confidence metrics (\ie reflective) should have an accuracy higher than standard purely probabilistic metrics (\ie intrinsic).
However, even reflective metrics rely on a form of probability rather than reflect any inherent internal characteristics of the model.
Therefore, if LAT can successfully capture internal representations of correctness, then we expect that LAT has an accuracy even higher than reflective metrics.

\subsubsection{Overall Performance}

\begin{table*}
\caption{\small LAT vs. baseline confidence metrics on \textbf{\textit{Test$_{\text{BCB}}$}} --- $^{*}$ indicates statistically significant improvement over all baselines ($p < 0.05$).\vspace{-0.3cm}}
\label{tab:metric_comparison_bcb}
\resizebox{\textwidth}{!}{
    \begin{tabular}{lrrrrrrrrrr}
    \toprule
    \multirow{2}{*}{\textbf{Model}} & \multirow{2}{*}{\textbf{Random}} & \textbf{Intrinsic} & \multicolumn{2}{c}{\textbf{Reflective}} & \multicolumn{3}{c}{\textbf{LAT (Val) -- Ours}} & \multicolumn{3}{c}{\textbf{LAT (Best)}} \\
    \cmidrule(lr){3-3}
    \cmidrule(lr){4-5}
    \cmidrule(lr){6-8}
    \cmidrule(lr){9-11}
    & & \textbf{Standard} & \textbf{Regular} & \textbf{True/False} & \textbf{\fit{BCB}} & \textbf{\fit{\mbppplus}} & \textbf{\fit{Syn}} & \textbf{\fit{BCB}} & \textbf{\fit{\mbppplus}} & \textbf{\fit{Syn}} \\
    \midrule
    Mistral & \multirow{4}{*}{27.5\%} & 5.7\% $\pm$ 0.4 & 28.9\% $\pm$ 0.8 & 23.7\% $\pm$ 0.0 & \textbf{51.1\% $\pm$ 10.6$^{*}$} & 47.8\% $\pm$ 0.8$^{*}$ & 33.0\% $\pm$ 0.1$^{*}$ & 57.9\% $\pm$ 4.5 & 59.3\% $\pm$ 0.9 & 48.4\% $\pm$ 0.5 \\
    Qwen & & 5.0\% $\pm$ 0.3 & 38.0\% $\pm$ 0.7 & 41.1\% $\pm$ 0.8 & \textbf{54.4\% $\pm$ 3.8$^{*}$} & 40.2\% $\pm$ 0.9 & 12.2\% $\pm$ 0.8 & 57.1\% $\pm$ 2.7 & 55.7\% $\pm$ 0.7 & 39.4\% $\pm$ 0.5 \\
    CodeLlama & & 4.8\% $\pm$ 0.4 & 24.7\% $\pm$ 0.4 & 23.7\% $\pm$ 0.0 & \textbf{50.3\% $\pm$ 6.2$^{*}$} & 25.8\% $\pm$ 0.3 & 22.2\% $\pm$ 0.7 & 55.0\% $\pm$ 3.0 & 48.2\% $\pm$ 0.9 & 39.0\% $\pm$ 0.7 \\
    OpenCoder & & 6.4\% $\pm$ 0.4 & 24.5\% $\pm$ 1.3 & 27.3\% $\pm$ 0.5 & \textbf{56.3\% $\pm$ 5.1$^{*}$} & 29.5\% $\pm$ 0.3 & 30.5\% $\pm$ 0.7 & 60.0\% $\pm$ 2.8 & 44.2\% $\pm$ 0.7 & 44.2\% $\pm$ 1.2 \\
    \bottomrule
    \end{tabular}
}
\end{table*}

\begin{table*}
\caption{\small LAT vs. baseline confidence metrics on \textbf{\textit{Test$_{\text{HE}}$}} --- $^{*}$ indicates statistically significant improvement over all baselines ($p < 0.05$).\vspace{-0.3cm}}
\label{tab:metric_comparison_he}
\resizebox{\textwidth}{!}{
    \begin{tabular}{lrrrrrrrrrr}
    \toprule
      \multirow{2}{*}{\textbf{Model}} & \multirow{2}{*}{\textbf{Random}} & \textbf{Intrinsic} & \multicolumn{2}{c}{\textbf{Reflective}} & \multicolumn{3}{c}{\textbf{LAT (Val) -- Ours}} & \multicolumn{3}{c}{\textbf{LAT (Best)}} \\
    \cmidrule(lr){3-3}
    \cmidrule(lr){4-5}
    \cmidrule(lr){6-8}
    \cmidrule(lr){9-11}
    & & \textbf{Standard} & \textbf{Regular} & \textbf{True/False} & \textbf{\fit{HE}} & \textbf{\fit{\mbppplus}} & \textbf{\fit{Syn}} & \textbf{\fit{HE}} & \textbf{\fit{\mbppplus}} & \textbf{\fit{Syn}} \\
    \midrule
    Mistral & \multirow{4}{*}{30.6\%} & 21.3\% $\pm$ 3.3 & 40.6\% $\pm$ 2.7 & 15.3\% $\pm$ 2.7 & \textbf{49.3\% $\pm$ 7.0$^{*}$} & 39.9\% $\pm$ 2.2 & 38.1\% $\pm$ 1.4 & 53.8\% $\pm$ 7.2 & 58.4\% $\pm$ 1.2 & 47.6\% $\pm$ 1.7 \\
    Qwen & & 36.8\% $\pm$ 3.1 & 7.0\% $\pm$ 1.1 & 6.7\% $\pm$ 0.9 & 60.7\% $\pm$ 8.3$^{*}$ & 39.6\% $\pm$ 1.3 & \textbf{63.0\% $\pm$ 2.0$^{*}$} & 67.8\% $\pm$ 4.6 & 65.6\% $\pm$ 1.8 & 65.2\% $\pm$ 1.9 \\
    CodeLlama & & 21.3\% $\pm$ 2.5 & 28.9\% $\pm$ 2.1 & 26.5\% $\pm$ 0.0 & \textbf{41.4\% $\pm$ 6.4$^{*}$} & 32.2\% $\pm$ 1.0 & 25.5\% $\pm$ 1.3 & 52.7\% $\pm$ 4.5 & 49.6\% $\pm$ 2.1 & 46.7\% $\pm$ 1.5 \\
    OpenCoder & & 36.5\% $\pm$ 3.9 & 31.0\% $\pm$ 2.5 & 9.9\% $\pm$ 1.6 & 44.0\% $\pm$ 8.3$^{*}$ & 35.1\% $\pm$ 1.8 & \textbf{47.8\% $\pm$ 3.0$^{*}$} & 56.5\% $\pm$ 3.1 & 52.4\% $\pm$ 2.8 & 56.6\% $\pm$ 3.9 \\
    \bottomrule
    \end{tabular}
}
\vspace{-0.2cm}
\end{table*}

We observe three patterns in Tables~\ref{tab:metric_comparison_bcb} and~\ref{tab:metric_comparison_he}. First, the intrinsic (standard) metric on \bcb\ hovers around 5--6\% for all models, whereas on \he\ it ranges from 21\%-37\%, reflecting the latter's lower difficulty. Second, reflective variants are inconsistent. For example, on \bcb, the true/false variant outperforms the regular variant for Qwen (41.1\% vs.\ 38.0\%) but underperforms for Mistral (23.7\% vs.\ 28.9\%). In contrast, on \he, Mistral's reflective regular accuracy (40.6\%) exceeds its true/false accuracy (15.3\%), while Qwen scores around 7\% on both. Third and most importantly, LAT consistently outperforms all baselines. On \bcb, it improves accuracy by +13.3 to +29.0 pp over the best reflective accuracy; on \he, gains range from +8.7 to +26.2 pp over the strongest baseline. These results show that capturing correctness from neural activity yields a more stable correctness signal than intrinsic and reflective confidence. Throughout the paper, we use Generalized Estimating Equations (GEE) logistic regression with Benjamini-Hochberg correction (p < 0.05) for significance testing.

\subsubsection{Out-of-Distribution Generalization}
We now analyze how well correctness representations transfer across datasets. On \bcb, fitting on \mbppplus yields between 25.8\% to 47.8\% accuracy with 3 out of 4 models above random but below \indistribution LAT (50.3\%--56.3\%). Synthetic fitting is erratic: two models drop below random (Qwen at 12.2\%, CodeLlama at 22.2\%), while Mistral and OpenCoder exceed it. On \he, \mbppplus fitting stays above random (32.2\%--39.9\%), but below in-distribution LAT (41.4--60.7\%). Synthetic fitting can match or exceed \indistribution LAT for some models (e.g., Qwen at 63.0\% vs.\ 60.7\%; OpenCoder at 47.8\% vs.\ 44.0\%), yet fails for others.
A different, yet related, observation is the standard deviation's pattern. \Indistribution fitting shows higher variance (e.g., $\pm 10.6\%$ under Mistral in \bcb) compared to the lower variance observed in \outofdistribution fitting (e.g., $\pm 0.8\%$ under Mistral in \bcb using \fit{\mbppplus}). This suggests a specialization versus generalization trade-off: fitting on \indistribution data achieves higher accuracy by specializing in data nuances (low bias), but this specialization leads to higher instability (high variance) when those nuances are missed across folds. On the other hand, fitting on \outofdistribution data captures more general features, despite leading to lower accuracy.

\subsubsection{Model-Specific Insights}
The magnitude of LAT's improvement compared to each model's strongest baseline reveals model differences. On \bcb, OpenCoder gains the most (+28.8 pp from 27.5\% to 56.3\%), followed by CodeLlama (+22.8 pp), Mistral (+22.2 pp), and Qwen (+13.3 pp). On \he, Qwen shows the largest increase (+26.2 pp from 36.8\% to 63.0\%), then OpenCoder (+11.3 pp), CodeLlama (+10.8 pp), and Mistral (+8.7 pp). These patterns suggest that while all models embed a correctness signal, its strength varies by training regime and task complexity. This is consistent with research highlighting that LLMs' ability to generalize, rather than merely memorize solutions, varies across models and different types of evolved coding tasks.~\cite{memorize_or_generalize}.

\subsubsection{Validation vs. Best Layer Selection}
We compare LAT (\textit{Val}), which selects layers via held-out validation, against LAT (\textit{Best}), the theoretical upper bound. Across models and datasets, the gap rarely exceeds 7pp for \indistribution fitting. On \bcb, Qwen scores 54.4\% (\textit{Val}) \vs 57.1\% (\textit{Best}) and, on \he, Mistral scores 49.3\% \vs 53.8\%. The greatest gap is OpenCoder with a difference of 12.5 pp between 44\% (\textit{Val}) and 56.5\% (\textit{Best}). Still, the proximity indicates that validation-driven layer choice reliably approximates the optimal layer without test-set leakage.
\vspace{-0.1cm}
\begin{tcolorbox}[colback=gray!10, colframe=black!50, top=1pt, bottom=1pt, title=RQ1: Key takeaway]
LAT successfully extracts meaningful correctness representations from neural activations, significantly outperforming random chance and surpassing intrinsic and reflective confidence metrics across both benchmarks and all models.
On \bcb, LAT improves accuracy by +13.3 to +28.8 pp over the best baseline; on \he, gains range from +8.7 to +26.2 pp.
\end{tcolorbox}

%% file: sections/rq2/rq2.tex
\section{RQ2: Can leveraging internal correctness representations lead to more effective correctness ranking?}
\label{sec:rq2}

Based on RQ1, we know that LAT can capture an internal notion of correctness that distinguishes correct from incorrect solutions. In RQ2, we explore using LAT to rank multiple generated solutions by estimated correctness. Traditionally, developers prompt LLMs for code, then evaluate outputs manually or with tests, often requiring several attempts. LAT can shortcut this by ranking candidates so the most correct are near the top, making it easier to select a correct implementation from a small set.

\input{sections/rq2/setup.tex}
\input{sections/rq2/results.tex}

%% file: sections/rq2/setup.tex
\subsection{Setup}
\label{sec:rq2-setup}

To evaluate LAT's ranking ability, we test whether it can effectively order multiple generations by correctness.

\subsubsection{Data Preparation}

We focus on both \he and \bcb, using the same tasks and reference solutions as RQ1 for fitting and validation in both \indistribution (\fit{HE}/\val{HE} and \fit{BCB}/\val{BCB}) and \outofdistribution (\fit{syn}\slash\val{syn} and \fit{\mbppplus}\slash\val{\mbppplus}). For testing, we use the same \testhetasks tasks from \test{HE} and \testbcbtasks tasks from \test{BCB} established in RQ1. For each task and model, we generate 10 diverse code samples (temperature = 1.0), yielding 1,210 samples per model for \he and 3,670 for \bcb. The goal is to assess each model's ability to rank its own outputs, similar to self-correction.

\subsubsection{LAT Setup}

We use the fit, validation, and test splits from the first fold of RQ1 to ensure consistency across models and tasks while simplifying the evaluation. The ranking process involves fitting correctness representations on the designated \fit{} set, selecting the optimal layer using the \val{} set, and applying the learned vector to score and rank the generated implementations in the \test{} set.

\subsubsection{Accuracy and Baselines}
We evaluate our ranking approach using pass@rank-$k$ metric, which measures the percentage of programming problems where at least one functionally correct solution (verified by test execution) appears within the top $k$ positions. Formally, for a set of problems $P$ and rank threshold $k$:
\vspace{-0.2em}
\[
\text{pass@rank-}k = \frac{1}{|P|} \sum_{p \in P} \mathbb{I}[\exists i \leq k : \text{correct}(\text{rank}_i(p))]
\]

where $\text{rank}_i(p)$ denotes the $i$-th ranked implementation for problem $p$, $\text{correct}(\cdot)$ indicates whether the implementation passes all unit tests, and $\mathbb{I}[\cdot]$ is the indicator function that returns 1 if the condition is true and 0 otherwise.

We report pass@rank-$k$ for $k \in \{1, \dots, 5\}$ on \he and \bcb to quantify the trade-off between the number of candidates and the probability that at least one in the top $k$ is correct. Note that unit tests are only used to evaluate the ranked outputs, not during the ranking process.
In addition to Random, Intrinsic, and Reflective metrics, we compare against these baselines: \vspace{-0.1cm}
\begin{itemize}[leftmargin=*]
    \item \textbf{pass@1 (baseline):} Fraction of problems where a single implementation passes all tests, representing single-attempt performance. We use temperature = 0.2 to balance greedy decoding (temperature = 0, which always selects the most likely token at each step) and diversity, helping avoid local minima while keeping solutions focused---in line with previous research~\cite{humaneval}. Note that this pass@1 baseline measures success using a single generation ($N=1$), while pass@rank-1 evaluates the top-ranked solution selected from the set of 10 diverse generations ($N=10$).
    \item \textbf{pass@10 (ceiling):} Fraction of problems where at least one of 10 implementations (same set used for pass@rank-$k$) passes all tests ($N=10$). We sample with temperature = 1, as this value should raise with the number of samples~\cite{alphacode}, to encourage diversity and maximize the chance of at least one correct solution, reflecting exploration rather than single-attempt success.
    \item \textbf{RankEF:} We also compare to RankEF~\cite{rankef}, a multi-task learning approach leveraging execution feedback to improve code ranking. During training, RankEF integrates classification labels and execution feedback to better distinguish correct from incorrect candidates. We report pass@rank-$k$ for RankEF on the same \bcb and \he test sets. Since no ready-to-use RankEF ranker was available, we reproduced the training process from the RankEF GitHub repository. Following the original paper, we used CodeT5+~\cite{codet5plus} as the base model, further trained on 5,000 APPS~\cite{apps-dataset} training tasks. For each task, we generated 100 samples with CodeT5+ and collected execution feedback for training. The model was trained using the described multi-task learning framework with hard parameter sharing. The authors report that training a CodeT5+-based RankEF model this way generalizes well and can rank generations from other models and datasets. Thus, we train a single RankEF model and use it to rank generations from all models in our experiments, where each model produces 10 candidate implementations per task. To validate our reproduction, we compared our results to the authors' reported scores on CodeLlama, the LLM both works have in common. Our reproduced ranker achieves slightly higher accuracy, likely because our evaluation ranks 10 candidate implementations per task, while the original work ranks 100.
\end{itemize}
\vspace{-0.37cm}

%% file: sections/rq2/results.tex
\subsection{Results}
\label{sec:rq2-results}

Figures \ref{fig:he_ranking_performance}-\ref{fig:bcb_ranking_performance} and Tables~\ref{tab:pass_rank1_he}-\ref{tab:pass_rank1_bcb} summarize the results of RQ2.
\vspace{-0.18cm}
\subsubsection{Overall Performance}
Our LAT-based ranking approach demonstrates significant improvements over the pass@1 baseline across both benchmarks, particularly as more candidates are considered ($k>1$). While simpler rankers like \texttt{Random} or \texttt{Intrinsic} show some gains, our LAT-based fittings (\fit{HE}/\fit{BCB}, \fit{\mbppplus}, \fit{Syn}) consistently provide a more robust path to higher accuracy. On the \he benchmark, this benefit is often immediate. For instance, Mistral's accuracy with \fit{HE} is 41.3\% at Rank-1, well above its 34.7\% pass@1 baseline and the 38.0\% achieved by \texttt{Random} selection---\autoref{fig:he_ranking_performance}. On the more challenging \bcb benchmark, where all models have lower baseline accuracies, the advantage of LAT-based ranking becomes even clearer at higher ranks, consistently outperforming the baselines and narrowing the gap to the pass@10 performance ceiling (See \autoref{fig:bcb_ranking_performance}). Compared to RankEF, an approach that needs a full fine-tune procedure on execution feedback, our LAT-based technique achieves competitive or superior accuracy without requiring expensive training or test execution, demonstrating the value of our lightweight ranking (see~\autoref{sec:discussion-trained-vs-fitted}). Numerically, LAT shows superior or comparable accuracy to all baselines, including RankEF, across most settings. Only a few comparisons reached statistical significance.

\subsubsection{Performance on \he}
\begin{figure}[t!]
    \centering
    \includegraphics[width=\columnwidth]{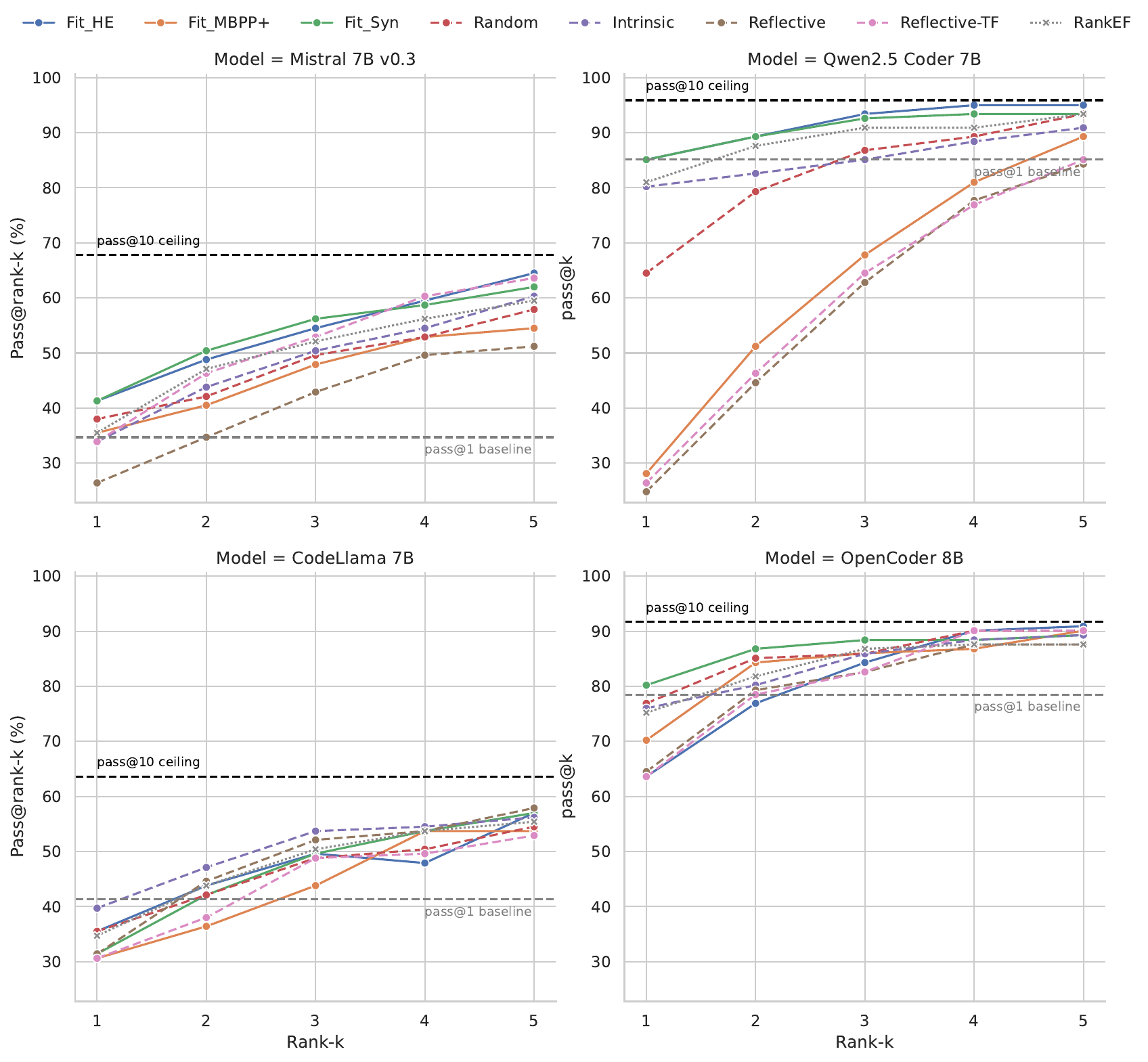}\vspace{-0.2cm}
    \caption{\small Accuracy of different ranking methods for \he compared to the pass@1 baseline and pass@10 ceiling.}
    \label{fig:he_ranking_performance}\vspace{-0.6cm}
\end{figure}

~\autoref{fig:he_ranking_performance} details the results for \he. The four panels highlight that while different models benefit from ranking to varying degrees, the LAT-based fittings are consistently effective.
For \textbf{Mistral}, the \fit{HE} and \fit{Syn} fittings provide the strongest performance, starting at 41.3\% and reaching 64.5\% and 62.0\% respectively at $k=5$, significantly outperforming all alternative methods and nearing the 67.8\% ceiling. Notably, both LAT fittings surpass RankEF, which in comparison achieved 35.5\% at $k=1$ and 59.5\% at $k=5$.
\textbf{Qwen}, with a very high 85.1\% pass@1 baseline, still benefits from LAT ranking. Its \fit{HE} and \fit{Syn} fittings match the baseline at $k=1$ and climb to 95.0\% and 93.4\% respectively by $k=5$, almost reaching the 95.9\% ceiling. This shows that even for highly capable models, LAT can effectively identify the best among several good candidates.
\textbf{CodeLlama} presents an interesting case where the \texttt{Intrinsic} metric is surprisingly effective, outperforming other methods at several ranks. This is likely due to the self-ranking context, where the model's generation probabilities are a strong ranking method, even though it depends on how well a particular model's output probabilities reflect actual correctness. However, our LAT-based fittings still show competitive performance, with \fit{HE} and \fit{Syn} reaching 57.0\% at $k=5$. Here, RankEF (34.7\% at $k=1$, 55.4\% at $k=5$) also shows competitive performance, though our LAT fittings remain higher.
For \textbf{OpenCoder}, \fit{Syn} and \fit{\mbppplus} provide the most substantial gains over its 78.5\% baseline, reaching 88.4\% and 86.0\% at $k=3$, respectively, demonstrating the power of out-of-distribution signals.
Across all models, the \texttt{Reflective} ranking methods consistently underperform, often doing worse than \texttt{Random}. This suggests that simple self-correction signals are unreliable, reinforcing the need for the more sophisticated capturing of correctness provided by LAT.

For easier reference, Table~\ref{tab:pass_rank1_he} summarizes the pass@rank-1 accuracy for all models and ranking methods on the \he benchmark. LAT-based fittings achieve the best rank-1 performance for Mistral (41.3\%), Qwen (85.1\%), and OpenCoder (80.2\%), while CodeLlama benefits most from its Intrinsic metric (39.7\%). Notably, our approach outperforms RankEF across all models.

\begin{table}
\caption{\small Comparison of Random, Intrinsic, Reflective, LAT, and RankEF pass@rank-1 accuracy (\%) on \he using \textbf{\textit{Test$_{\text{HE}}$}}.\vspace{-0.4cm}}
\label{tab:pass_rank1_he}
\resizebox{\columnwidth}{!}{
    \begin{tabular}{lcccccccccc}
    \toprule
    \multirow{2}{*}{\textbf{Model}} & \multirow{2}{*}{\textbf{Random}} & \textbf{Intrinsic} & \multicolumn{2}{c}{\textbf{Reflective}} & \multicolumn{3}{c}{\textbf{LAT (Val) -- Ours}} & \multirow{2}{*}{\textbf{RankEF}} \\
    \cmidrule(lr){3-3}
    \cmidrule(lr){4-5}
    \cmidrule(lr){6-8}
    & & \textbf{Standard} & \textbf{Regular} & \textbf{T/F} & \textbf{\fit{HE}} & \textbf{\fit{MBPP}} & \textbf{\fit{Syn}} & \\
    \midrule
    Mistral    & 38.0\% & 33.9\% & 26.4\% & 33.9\% & \textbf{41.3\%} & 35.5\% & \textbf{41.3\%} & 35.5\% \\
    Qwen       & 64.5\% & 80.2\% & 24.8\% & 26.4\% & \textbf{85.1\%} & 28.1\% & \textbf{85.1\%} & 81.0\% \\
    CodeLlama  & 35.5\% & \textbf{39.7\%} & 31.4\% & 30.6\% & 35.5\% & 30.6\% & 31.4\% & 34.7\% \\
    OpenCoder  & 76.9\% & 76.0\% & 64.5\% & 63.6\% & 63.6\% & 70.2\% & \textbf{80.2\%} & 75.2\% \\
    \bottomrule
    \end{tabular}
}
\end{table}

\vspace{-0.2cm}
\subsubsection{Performance on \bcb}

\begin{figure}[t!]
    \centering
    \includegraphics[width=\columnwidth]{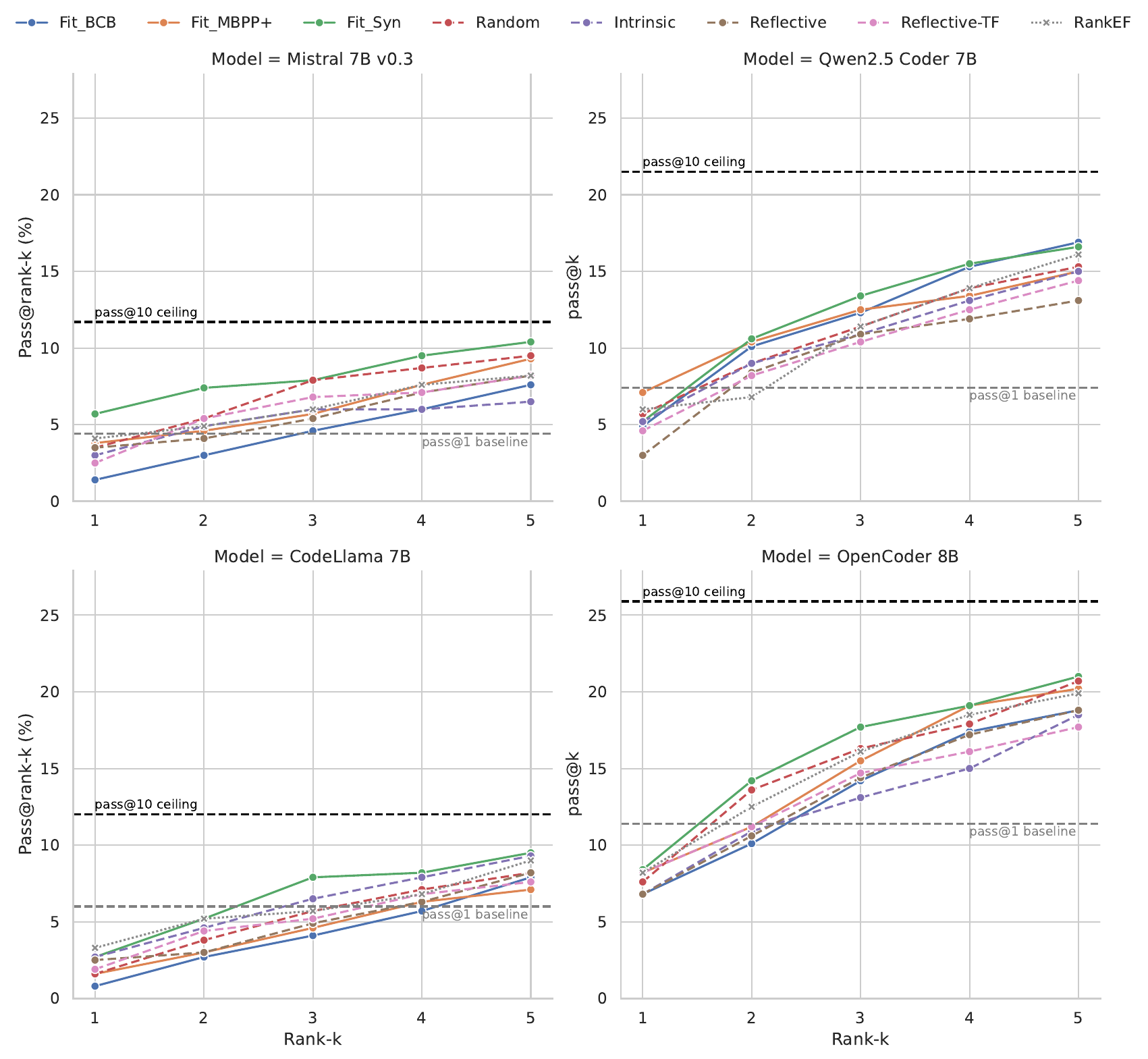}\vspace{-0.2cm}
    \caption{\small Accuracy of different ranking methods for \bcb compared to the pass@1 baseline and pass@10 ceiling.}
    \label{fig:bcb_ranking_performance}\vspace{-0.5cm}
\end{figure}

\autoref{fig:bcb_ranking_performance} shows the results on the more difficult \bcb benchmark.
For \textbf{Mistral}, the \fit{Syn} fitting is the most effective, improving from 5.7\% at $k=1$ to 10.4\% at $k=5$, consistently staying above all other methods. \fit{Syn} already outperforms RankEF at rank-1 (5.7\% vs. 4.1\%) and maintains a clear lead across all ranks, reaching 10.4\% at rank-5 compared to RankEF's 8.2\%. In contrast, we find that by rank-2, RankEF does not outperform the Random baseline.
\textbf{Qwen} sees all three LAT-based fittings surpass the baselines and alternative methods by $k=2$, with \fit{BCB} and \fit{Syn} reaching 16.9\% and 16.6\% at $k=5$.
\textbf{CodeLlama} struggles on this benchmark, only able to surpass the pass@1 baseline and non-LAT methods by rank $k=3$. Here, RankEF performs best at $k=1$ (3.3\%), suggesting that execution feedback may be particularly valuable when models struggle with task difficulty. However, despite the lower start at rank-1 (2.7\%) and tying RankEF at rank-2, \fit{Syn} provides the highest accuracy for the remaining ranks, achieving 9.5\% at $k=5$.
\textbf{OpenCoder} shows significant gains with out-of-distribution fittings. The \fit{Syn} curve rises sharply with \fit{\mbppplus} catching on by rank $k=4$, eventually reaching 21.0\% and 20.2\% at $k=5$, surpassing the 11.4\% baseline and demonstrating that LAT-based ranking is more effective than non-LAT methods.
On this benchmark, the \texttt{Intrinsic} metric is far less effective than on \he, highlighting its unreliability as a general strategy. In contrast, the LAT-based fittings, particularly \fit{Syn}, prove to be a consistently effective method for improving performance across all models on \bcb.

Table~\ref{tab:pass_rank1_bcb} summarizes the pass@rank-1 on the more challenging \bcb benchmark. Overall, LAT's \fit{Syn} fitting achieves top performance for Mistral (5.7\%) and OpenCoder (8.4\%) while LAT's \fit{MBPP} achieves the highest score for Qwen (7.1\%). With the exception of CodeLlama, LAT-based ranking achieves higher accuracy than RankEF for all models at rank-1.

\begin{table}
\caption{\small Comparison of Random, Intrinsic, Reflective, LAT, and RankEF pass@rank-1 accuracy (\%) on \bcb using \textbf{\textit{Test$_{\text{BCB}}$}}.\vspace{-0.4cm}}
\label{tab:pass_rank1_bcb}
\resizebox{\columnwidth}{!}{
    \begin{tabular}{lcccccccc}
    \toprule
    \multirow{2}{*}{\textbf{Model}} & \multirow{2}{*}{\textbf{Random}} & \textbf{Intrinsic} & \multicolumn{2}{c}{\textbf{Reflective}} & \multicolumn{3}{c}{\textbf{LAT (Val) -- Ours}} & \multirow{2}{*}{\textbf{RankEF}} \\
    \cmidrule(lr){3-3}
    \cmidrule(lr){4-5}
    \cmidrule(lr){6-8}
    & & \textbf{Standard} & \textbf{Regular} & \textbf{T/F} & \textbf{\fit{BCB}} & \textbf{\fit{MBPP}} & \textbf{\fit{Syn}} & \\
    \midrule
    Mistral    & 3.5\% & 3.0\% & 3.5\% & 2.5\% & 1.4\% & 3.8\% & \textbf{5.7\%} & 4.1\% \\
Qwen       & 5.7\% & 5.2\% & 3.0\% & 4.6\% & 4.9\% & \textbf{7.1\%} & 5.2\% & 6.0\% \\
CodeLlama  & 1.6\% & 2.7\% & 2.5\% & 1.9\% & 0.8\% & 1.6\% & 2.7\% & \textbf{3.3\%} \\
OpenCoder  & 7.6\% & 6.8\% & 6.8\% & 8.2\% & 6.8\% & 8.2\% & \textbf{8.4\%} & 8.2\% \\
    \bottomrule
    \end{tabular}
}
\end{table}

\subsubsection{Out-of-distribution Generalization}
A key strength of our approach is its ability to generalize. On \he, although the \indistribution \fit{HE} fitting often leads, the \outofdistribution \fit{Syn} fitting matches or exceeds it at low $k$ for Mistral and Qwen. This contrasts with our MCQA setup (RQ1), where incorrect options came from other models and \outofdistribution fittings underperformed. In RQ2, all candidates are self-generated, so a model's internal correctness signal may align more consistently. For OpenCoder, \fit{Syn} achieves 80.2\% at $k=1$, versus 63.6\% for \fit{HE}. Notably, the \outofdistribution \fit{Syn}'s generalization outperforms RankEF for all ranks despite RankEF's more specialized training, highlighting the transferability of LAT correctness patterns.

On \bcb, \outofdistribution fittings (\fit{Syn}, \fit{\mbppplus}) frequently outperform \fit{BCB}. For Mistral and OpenCoder, \fit{Syn} is top across almost all ranks. This robust transfer suggests a single well-fitted LAT can rank solutions on new tasks without any \indistribution data and often better than an \indistribution fit.

\subsubsection{Rank Threshold Analysis}

Across both benchmarks, accuracy consistently improves as the rank threshold $k$ increases from 1 to 5, demonstrating that the LAT-based method effectively favors correct solutions from the candidate pool. While the top-ranked solution ($k=1$) is not always correct, expanding consideration to the top 3 or 5 candidates captures most of the achievable performance, rapidly approaching the pass@10 ceiling. For example, on \he, Mistral improves from its 34.7\% pass@1 baseline to 64.5\% at $k=5$ with \fit{HE}, closing most of the gap to its 67.8\% ceiling. A similar pattern holds on \bcb, where OpenCoder jumps from an 11.4\% baseline to 21.0\% at $k=5$ with \fit{Syn}. Comparing growth trajectories between LAT and RankEF reveals model-dependent patterns. For Mistral, LAT's \fit{Syn} starts higher at rank-1 (5.7\% vs. RankEF's 4.1\%) and maintains its lead through rank-5 (10.4\% vs. 8.2\%). For Qwen, \fit{BCB} and \fit{Syn} start lower at rank-1 (4.9\% and 5.2\% vs. RankEF's 6.0\%) but overtake RankEF by rank-2 (10.1\% and 10.6\% vs. RankEF's 6.8\%), showing steeper improvement. For CodeLlama, LAT's \fit{Syn} ties RankEF at rank-2 (5.2\%) and then surpasses it at rank-3 onwards, reaching 9.5\% at rank-5 versus RankEF's 9.0\%. For OpenCoder, LAT's \fit{Syn} shows competitive or superior performance across all ranks. Overall, LAT fittings demonstrate a pattern of rapid early improvement (rank-1 to rank-3), effectively catching up to and surpassing RankEF even when off to a slightly lower start for CodeLlama. 

\vspace{-0.2cm}

\subsubsection{Model-Specific Insights}
On \he, models with lower baselines like Mistral see the largest relative gains (an 85.9\% relative improvement from pass@1 to $k=5$ with \fit{HE}). For high-performing models like Qwen, LAT is still able to push it towards its ceiling. 
RankEF also shows consistent improvements over the pass@1 baseline, but LAT's best fittings achieve comparable or superior gains. For Mistral, LAT's \fit{HE} achieves +29.8 points over pass@1 (vs. RankEF's +24.8), while for OpenCoder, \fit{HE} achieves +12.4 (vs. RankEF's +9.1). For CodeLlama, LAT's gains with \fit{Syn} are +15.7 points, compared to RankEF with +14.1 points.

On the more challenging \bcb benchmark, these model-dependent patterns highlight the robustness of our ranking approach. Qwen improves its accuracy by 9.5\% absolute points (from 7.4\% to 16.9\% at $k=5$ with \fit{BCB}), while OpenCoder achieves a 9.6\% absolute gain (from 11.4\% to 21.0\% with \fit{Syn}). The consistent success of the \fit{Syn} fitting, especially on \bcb, suggests that training on a diverse, synthetic dataset equips the LAT model with a strong and generalizable understanding of code correctness.

\begin{tcolorbox}[colback=gray!10, colframe=black!50, top=1pt, bottom=1pt, title=RQ2: Key takeaway]
LAT-based ranking outperforms pass@1 baseline, intrinsic, and reflective methods on \he and \bcb across all models. LAT showed significant gains by rank $k=3$ and further closed most of the gap to the pass@10 ceiling by rank $k=5$. 
Except for CodeLlama, LAT achieved higher accuracy than RankEF for all models at rank-1.
A key finding is the effectiveness of \outofdistribution fittings (\fit{Syn} and \fit{\mbppplus}), which highlights LAT's potential to reduce costly test executions and generalize to new tasks without requiring \indistribution data.
\end{tcolorbox}
\vspace{-0.3cm}

%% file: sections/discussion.tex
\section{Discussion}
\label{sec:discussion}
We now discuss the implications of our findings in terms of practical utility, differences between our approach and training-based ones, as well as directions for future research.
\vspace{-0.15cm}
\subsection{Practical Utility}
The ability to capture internal correctness representations has practical applications for developers. Our ranking method in RQ2 shows that these representations can be used to select promising code solutions without running tests~\cite{neural_code_rankers}. In essence, when an LLM generates multiple code solutions, our method can filter out incorrect candidates and highlight promising ones. Thus, developers can focus their verification efforts on a smaller set of top-ranked solutions, rather than having to examine every candidate. 
It could even be used ``behind the scenes'' where the LLM-based generation system internally generates multiple candidates and returns only the highest ranked candidate to the developer, effectively increasing the accuracy of the single generation the developer sees.
Beyond this, our technique can be integrated into the software development life cycle. Since our approach captures relative correctness, it is best suited for comparing code changes. For example, in a CI/CD pipeline, it can flag changes where the new code appears less correct than the previous version, and prioritize test cases based on which code is deemed less correct.~\cite{survey_testing_minimization} In IDEs, it can provide confidence scores for code suggestions relative to the current implementation, or compare multiple suggestions against each other.~\cite{prompter_self_confident_recommender}.
\vspace{-0.25cm}
\subsection{Fitted Representations vs. Trained Rankers}
\label{sec:discussion-trained-vs-fitted}
Our accuracy comparisons with RankEF showed superior or comparable results. We discuss some additional practical advantages that our LAT-based ranking approach offers over trained rankers.

\noindent\textbf{Setup cost and data efficiency.} RankEF requires training a separate ranker model on thousands of code samples, which is memory- and compute-intensive. Moreover, gathering execution feedback for RankEF's training dataset requires running test suites on thousands of generated samples, adding substantial overhead to the setup. Reproducing RankEF involved 72 GPU-hours for sample generation, 10 CPU-hours for execution feedback collection, and 90 GPU-hours for model training on an NVIDIA A100 (peak memory: 60GB). Our technique, by contrast, is \emph{fitted} rather than \emph{trained}---requiring only PCA fitting on a small dataset. Average fitting time (PCA + Validation) is 3.75 seconds. This is computationally inexpensive and data-efficient, which is especially evident from our synthetic fitting set of only of 5 triplets of tasks with correct and incorrect implementations that fits in around 1 second.

\noindent\textbf{Flexibility and adaptability.} The lightweight fitting process makes our approach more adaptable. Should correctness criteria evolve (e.g., from `correctness' to `readability' or `adherence to new style guidelines'), our technique can be re-fitted on a small set of relevant examples. Training-heavy approaches like RankEF would require expensive re-training with large datasets.

\noindent\textbf{Inference and generalizability trade-offs.} Our technique requires extracting hidden states during generation, resulting in higher per-task inference time (0.399s vs. RankEF's 0.202s). However, this $\sim$0.2s overhead per task is negligible compared to the amortized setup cost: RankEF's 172 GPU+CPU-hours would need to rank roughly 3 million tasks to break even on compute cost alone. For practical scenarios, LAT's lightweight setup far outweighs its inference overhead. Additionally, while our approach is model-specific (representations captured from one LLM cannot be applied to another's hidden states), a model and its fitting can rank code solutions from any source. RQ1 demonstrates this, where a choice was made on candidates from different sources: reference solution and other LLMs (see~\autoref{sec:rq1-data-prep}).
\vspace{-0.25cm}
\subsection{The Nature of Internal Correctness}
Our findings strongly indicate that LLMs develop an internal representation of code correctness---in line with similar observations in reasoning models~\cite{reasoning_models_know_when_theyre_right}. More concretely, the core finding is that correctness representations extracted via RepE outperform baseline confidence metrics~\cite{assessing_correctness_via_uncertainty_estimation}. In RQ1, LAT improves accuracy over reflective baselines on both \bcb and \he. In RQ2, LAT-based ranking improves direct pass@1 performance and approaches the pass@10 ceiling. Overall, this indicates that a correctness signal captured from the model's internal states provides a more stable indicator of correctness than the model's output probabilities, which are poorly calibrated with correctness~\cite{calibration-and-correctness-icse25}.
In this work, we showed that this signal \textit{exists and can be leveraged} but not \textit{why or how} it exists. Is the model learning a deep semantic understanding of the code's logic? Or is it learning more superficial pattern recognition based on syntactic structures and token sequences that correlate with correctness in the training data? It would be interesting to explore this distinction in future work.
\vspace{-0.2cm}
\subsection{Beyond Functional Correctness}
In this paper, we focused on functional correctness, which can be objectively verified using tests.
Since the represented concept depends on the contrasting stimuli presented to the LLM, future work could explore whether LAT can also capture other non-functional properties such as maintainability or efficiency~\cite{beyond_functional_correctness, unveiling_inefficiencies}.
\vspace{-0.2cm}

%% file: sections/t2v.tex
\section{Threats to Validity}
\label{sec:t2v}
\textbf{Construct Validity.} We use passing unit tests as a proxy for correctness, while tests may fall short when it comes to non-functional aspects (\eg performance and readability). There are two factors that can affect the correctness representation captured by LAT: the stimuli used and the model layer chosen to capture the correctness signal.
Since our primary goal in this paper is to rigorously explore if RepE can be used for code correctness, we carefully adhered to the same stimulus setup and layer selection used in the original work to ensure methodological consistency. 
Specifically, we employ a robust selection method that verifies which layer has the highest accuracy on a validation split. We also evaluate on two different benchmarks (\he and \bcb), combining ID and OOD stimuli, and doing CV.

\textbf{Internal Validity.} Prompt formatting choices or verbalized confidence levels could influence our observations.
We reduce prompt bias by using prompt designs that are consistent with prior work.
The sourcing of candidate solutions also differs between RQ1---where incorrect options come from other LLMs---and RQ2---where variants are self-generated---which may partly explain divergent \outofdistribution behavior. Evaluating both settings exposes how candidate origin impacts performance.
Also, collecting incorrect implementations from existing LLMs could introduce bias if those models share training data or produce similar errors to the target model. 
We address this by collecting failing solutions from multiple reputable models and using an established evaluation harness~\cite{bigcode_evaluation_harness} and benchmarks (\he and \bcb)

\textbf{External Validity.} We evaluate the accuracy of four 7-8B parameter LLMs on two Python benchmarks (\he and \bcb). 
To improve generalization, we include both general and code models, evaluate on benchmarks with differing degrees of complexity, and test LAT on OOD data (synthetic and \mbppplus).
However, the observed LAT effectiveness and ranking improvements may not hold for larger or smaller models, different architectures, or other programming languages.
Our single-function evaluation does not cover some aspects of real-world code, \eg multi-file projects or integration tests, potentially found in repository-level benchmarks such as SWE-Bench~\cite{swebench}. We see our work not as a direct tool for such settings, but as a component within an LLM-based agent's strategy. We plan to explore such settings as future work.
\vspace{-0.2cm}

%% file: sections/related-work.tex
\section{Related Work}
\label{sec:related-work}
LLMs have improved developer productivity and efficiency~\cite{llm_systematic_literature_review}, but concerns remain about the quality of generated code, including incorrectness~\cite{understanding_code_generation_errors} and security vulnerabilities~\cite{asleep_at_the_keyboard}.

Recent work has explored training neural rankers to select code solutions without test execution. CodeRanker~\cite{neural_code_rankers} uses fault-aware classification to predict program correctness. RankEF~\cite{rankef} improves upon this with multi-task learning on execution feedback, enabling it to learn failure causes.

Many studies focus on estimating LLM confidence and improving calibration. Traditional methods often use output probabilities~\cite{measuring_massive_multitask}, few-shot prompting~\cite{a_few_more_examples}, or verbalized confidence~\cite{tian-etal-2023-just}. However, neural networks are often miscalibrated~\cite{calibration_modern_neural_networks}, and standard metrics may not reliably reflect correctness in code generation~\cite{calibration-and-correctness-icse25,dont_judge_code_by_its_cover}, with added concerns in cyber threat intelligence~\cite{llms_cyber_threat_intelligence}.

To address these issues, recent work explores more sophisticated confidence metrics. One approach infers confidence from the consistency of multiple generated outputs~\cite{showing_llm_code_selectively}, linking program similarity to correctness and reducing errors. Similar consistency-based methods have been applied to NL tasks~\cite{luq_long_text_uq}.

LLM internal states have also been leveraged for confidence estimation, such as through contrastive learning across layers~\cite{internal_inspector} or by weighting tokens using attention values~\cite{contextualized_sequence_likelihood}, reinforcing the potential of internal signals for output quality.

AI interpretability and transparency are increasingly important~\cite{machines_humans_focus_code,rethinking_interpretability}. Representation engineering~\cite{rep-eng} systematically studies how concepts are encoded in hidden states and has been used to detect and control issues like dishonesty in NL.

Bui \etal~\cite{correctness_assessment_llms_internal_representations} also use LLM internal states for code correctness, but classify single programs, while we rank candidates. Our method uses RepE to capture differences between correct and incorrect programs, unlike their separate classifier approach.

Huang \etal~\cite{risk_assessment_internal_states} use internal states for line-level risk assessment, flagging potentially incorrect code. Their classification-based setup (on tasks like code repair, translation, and editing) differs from our ranking approach and program synthesis benchmarks we used.

%% file: sections/conclusion.tex
\section{Conclusion}
\label{sec:conclusion}

In this work, we demonstrated that LLMs encode meaningful internal representations of code correctness. We adapted RepE to source code, validated its effectiveness on \he and \bcb, and demonstrated that our representation-based selection method outperforms basic output probabilities and reflective confidence metrics. We extended our efforts by implementing a ranking setup that leverages these correctness signals to select higher-quality solutions from multiple generations, improving pass@1 performance and approaching a theoretical pass@10 ceiling. We also achieve superior or comparable performance to the state-of-the-art RankEF ranker model across most settings. The data and code necessary to replicate our experiments is publicly available.~\cite{replication_package} We consider continued exploration of LLM internals promises further improvements in trustworthy AI-assisted programming.